\documentclass[trackchanges]{aastex701}

\usepackage{gensymb}
\usepackage{rotating}
\usepackage{lineno}
\usepackage{hyperref}
\usepackage{caption} % Required for captions
\usepackage{subcaption}
\usepackage{amsmath}

\begin{document}

\title{Tracking the Activity of the Interstellar Object 3I/ATLAS through its Perihelion}

\correspondingauthor{T. Marshall Eubanks}

\author[0000-0001-9543-0414]{T. Marshall Eubanks}
\affiliation{Space Initiatives Inc, Princeton, WV 24740, USA}
\email[show]{tme@space-initiatives.com}

\author[0000-0002-7164-2786]{Craig E. DeForest}
\affiliation{Southwest Research Institute, Boulder, CO, USA}
\email{craig.deforest@swri.org}

\author[0000-0002-0906-1761]{Kevin J. Walsh}
\affiliation{Southwest Research Institute, Boulder, CO, USA}
\email{kevin.walsh@swri.org}

\author[0000-0003-0333-6055]{Simon Porter}
\affiliation{Southwest Research Institute, Boulder, CO, USA}
\email{porter@swri.org}

\author[0009-0004-3836-8166]{Thomas Lehmann}
\affiliation{School Observatory Friedrich-Schiller-Gymnasium, Weimar, Germany}
\email{t.lehmann@mailbox.org}

\author[0000-0002-1978-8243]{Bruce G. Bills}
\affiliation{Jet Propulsion Laboratory, 4800 Oak Grove Drive, Pasadena, CA 91109, USA}
\email{Bruce.Bills@jpl.nasa.gov}

\author[0000-0003-1116-576X]{Adam Hibberd}
\affiliation{Institute for Interstellar Studies (i4is US), Oak Ridge, TN 37830, USA}
\email{hibberdadam994@gmail.com}

\author{W. Paul Blase}
\affiliation{Space Initiatives Inc, Princeton, WV 24740, USA}
\email{wpb@space-initiatives.com}

\author[0000-0003-1763-6892]{Andreas M. Hein}
\affiliation{SnT, University of Luxembourg,  L-4365 Luxembourg}
\email{andreas.hein@uni.org}
\affiliation{Initiative for Interstellar Studies, London,  SW8 1SZ, UK}
\email{andreas.hein@i4is.org}

\author[0000-0003-3924-7935]{Robert G. Kennedy III}
\affiliation{Institute for Interstellar Studies (i4is US), Oak Ridge, TN 37830, USA}
\email{robert.kennedy@i4is.org}

\author[0000-0003-1322-7485]{Adrien Coffinet}
\affiliation{Coulommiers, France}
\email{adrien.coffinet2@gmail.com}

\author[0000-0003-0626-1749]{Pierre Kervella}
\affiliation{LIRA, Observatoire de Paris, Universit\'e PSL, Sorbonne Universit\'e, Universit\'e Paris Cit\'e, CY Cergy Paris Universit\'e, CNRS, 5 place Jules Janssen, 92195 Meudon, France}
\affiliation{French-Chilean Laboratory for Astronomy, IRL 3386, CNRS and U. de Chile, Casilla 36-D, Santiago, Chile}
\email{pierre.kervella@obspm.fr}
\author[0009-0007-3343-2001]{Carlos Gomez de Olea Ballester}
\affiliation{TU Munich, D-80333 Munich, Germany}
\email{carlos.olea@tum.de}
%% The 1,500 word count limit includes title, headers, captions, and references with 150 words reserved for the required abstract.
\begin{abstract}
In order to facilitate interplanetary spacecraft observations of 3I/ATLAS, we have monitored and predicted the optical properties of its coma using both ground and space-based observations. Here, we describe how the data from 
space-based solar coronagraphs and the PUNCH mission  enabled tracking of 3I/ATLAS’s  optical magnitude throughout its entire perihelion passage,  including the period between October 8 and 30, 2025, when it was not visible from Earth. 
\end{abstract}
\keywords{\uat{Exocomets}{2368} -- \uat{Interstellar objects}{52}  --- \uat{Comet dynamics}{2213}}

%% That way, if you change tXiong’anhe order of any elements, LaTeX willM81twitter--AsteroidEnergy

\section{Introduction} 

The trajectory of 
3I/ATLAS passes relatively close to several spacecraft, allowing imaging from several spacecraft at Mars, and the Psyche, Juice and Juno interplanetary spacecraft \citep{Eubanks-et-al-2025-a}. 
Spacecraft observations of a target of opportunity such as 3I/ATLAS  require considerable planning, creating a need for advance predictions of the target's optical properties. We have been providing such predictions by modeling the magnitude drift  of the 3I/ATLAS coma, and also by obtaining and analyzing  a wide variety of ground and space based optical data.

\section{Optical Observations of 3I/ATLAS}
\label{sec:observations}

Figure \ref{fig:Magnitude-Data} shows 
optical magnitudes for 3I/ATLAS from a variety of sources.
Comet coma magnitude estimates are highly sensitive to aperture and color, as their diffuse edges mean wider fields of view collect more light, increasing flux and lowering magnitude. Additionally, gas comas have strong  spectral emissions,  which can bias magnitudes depending on the pass-band. We initially focused on G band as a proxy for the pass-bands of spacecraft telescope cameras.

At present the Minor Planet Center Database (MPCD)
contains 3989 entries for 3I/ATLAS in 14 different spectral bands, with 48.8\% of these entries being in the broad (400 to 1000 nm) G band displayed in Figure \ref{fig:Magnitude-Data}. 
In  early September, 2025, it became clear that 3I/ATLAS had developed a gas coma, with a characteristic green color from C$_{2}$ emissions.
Since July 2nd, coauthor
Thomas Lehmann has acquired a long series of green-filtered optical images and magnitudes, shown on Figure \ref{fig:Magnitude-Data} as green squares, in order to obtain visual equivalent magnitudes   using a green filter 
or the green color channel to emphasize gas
coma light over dust. The imaging  reduction pipeline was also intended to detect gas coma light to a larger angular extent than typical imaging. These data match the MPCD G band measurements well up until early September, and then clearly show a faster magnitude drift rate.

The Comet Observation Database (COBS), maintained by Crni Vrh Observatory and open to amateur contributions, records coma and tail sizes. For 
3I/ATLAS, data points with coma diameter $>$1' (red dots in Figure \ref{fig:Magnitude-Data})
match the trend of the Lehmann Green Filter magnitudes well with a bias of
~+0.5 magnitudes.

\subsection{The Polarimeter to Unify the Corona and Heliosphere}
\label{subsec:punch}

The Polarimeter to Unify the Corona and Heliosphere (PUNCH) images the solar corona and solar wind   from low-Earth orbit \citep{DeForest-et-al-2025-a}.   PUNCH observes  solar elongations from $1.25^\circ$ to $45^\circ$ in 450-750 nm broadband visible light, with a spatial resolution of $\sim$2 arcmin.  

\begin{figure}[ht!]
\centering
  \includegraphics[clip,width=\columnwidth]{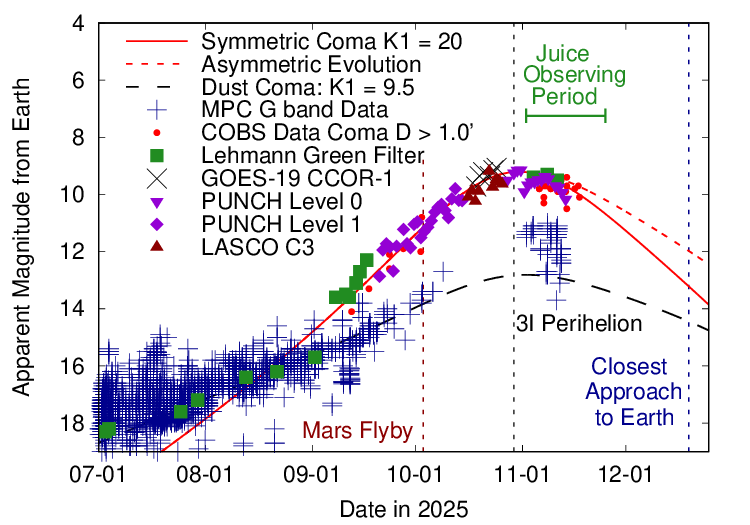}%
  \caption{3I/ATLAS magnitude data from a variety of sources, together with the Nov 6th  prediction models in Equations \ref{eq:coma-parameters-gas} and \ref{eq:coma-parameters-dust}.}
    \label{fig:Magnitude-Data}
\end{figure}

The PUNCH Wide Field Imager (WFI) uses conventional optics and deep baffles to attenuate stray light and observes for solar elongations between  45$\degree$ and 5$\degree$. 
Four spacecraft work together synchronously to form a single ``virtual coronagraph’’ that collects seven deep-field exposures every eight minutes, including one through a clear aperture. 
The clear images are stacked for an entire day, typically 174 to 176 images, and
are shown as purple diamonds and triangles in  Figure \ref{fig:Magnitude-Data}.  (The PUNCH levels refer to the  amount of image processing in the PUNCH data analysis.) At lower elongations PUNCH observes with its Narrow Field Imager (NFI); these data are still being processed.

%% The "ht!" tells LaTeX to put the figure "here" first, at the "top" next
%% and to override the normal way of calculating a float position.
%% The asterisk after "figure" tells the compiler to span multiple columns
%% if a two column style is selected.

\subsection{Solar Coronagraphs}
\label{subsubsec:Coronagraphs}

When 3I gets very close to the Sun as seen from the Earth, it could still be observed by orbiting solar coronagraphs, telescopes which block the light of the Sun  \citep{Zhang-Battams-2025-a}. Coronagraph estimates of 3I/ATLAS magnitudes are available from  the C3 camera of the 
Large Angle and Spectrometric COronagraph (LASCO) instrument of the SOHO spacecraft from 2025 Oct 16 – 26 (brown triangles in Figure \ref{fig:Magnitude-Data}) and from the CCOR-1 coronagraph on the NOAA GOES-19 satellite from Oct 9 through Nov 1 (black crosses in Figure \ref{fig:Magnitude-Data}).  These magnitudes are all based on the stacking of images over an entire day \citep{Zhang-Battams-2025-a}.

\section{Prediction of the 3I/ATLAS Magnitude}
\label{sec:Magnitudes}
 
Predictions for comet magnitudes, m, can be provided by a simple empirical equation
\citep{Shanklin-2023-a}
\begin{equation}
\mathrm{m}\ =\ \mathrm{H}_{1}\ + 5\ 
\log_{10}(\mathrm{D}_{\mathrm{observer}})\ +\ 
\mathrm{K}_{1}\ \log_{10}(\mathrm{R}_{\mathrm{Sun}})
\label{eq:comet_magnitudes}
\end{equation}
%% comet ma
where H$_{1}$ is the absolute magnitude,  K$_{1}$ 
is the magnitude drift rate,
D$_{\mathrm{observer}}$ is the distance (in Astronomical Units, or au) between the observer and the comet,  and R$_{\mathrm{Sun}}$ is the distance (in au) between the comet and the Sun. Due to a lack of suitable data, we do not  include a model for magnitude changes with phase angles.

After inquiries from the MRO HiRISE team we started directly 
determining these parameters in September, 2025, which led to the Sep 18th model, which used a K$_{1}$ of 20. 
This model worked reasonably well until PUNCH data for October became available, when it became clear  the magnitude model needed updating,  first on October 28th, and then again on Nov 6th with the initial post-perihelion data. 

The Nov 6th model also implemented the asymmetric magnitude drift model in \cite{Lacerda-et-al-2025-a}, based on fits to 92 comets with both pre and post perihelion magnitude data. In this model 
\begin{equation}
\mathrm{K}_{2}\ -\ \mathrm{K}_{1}\ =\ 
-0.65\ \times\ ( \mathrm{K}_{1}\ -\ 11.6)
\label{eq:assymmetric-model}
\end{equation} 
where K$_{2}$ is the post-perihelion drift rate.
The November 6th model is, for the gas coma 
\begin{equation}
\begin{split}
\mathrm{K_{1}} = 20.0 \\
\mathrm{H_{1}} = 4.75				    \\
\mathrm{K_{2}} = 14.54 \\
\mathrm{H_{2}} =  5.48 \\
\end{split}
\label{eq:coma-parameters-gas}
\end{equation}
where H$_{2}$ is the post-perihelion absolute magnitude (this bias must change to keep the model curve continuous).
The symmetric model simply uses the model H$_{1}$ and K$_{1}$ in the post-perihelion period. 
We decided there was not enough information to adjust the dust coma model, and did not change it from the Oct 28th version,
\begin{equation}
\begin{split}
\mathrm{K_{1}} = 9.75\\
\mathrm{H_{1}} = 9.5				    \\
\end{split}
\label{eq:coma-parameters-dust}
\end{equation}
Both of these models are shown in Figure \ref{fig:Magnitude-Data}, which also includes the symmetric version  of Equation \ref{eq:coma-parameters-gas}. It is clear that more observations will be needed to directly determine the post-perihelion K$_{2}$ from the data. 

\section{Conclusions} \label{sec:conclusions}

Spacecraft observations have enabled continuous monitoring of 3I ATLAS throughout its perihelion period, providing information and predictions for  interplanetary spacecraft observations. It is also clear that there was no substantial splitting of the 3I/ATLAS nucleus or other unusual behavior during its perihelion period. 

We intend to continue 3I/ATLAS monitoring and analysis through the close approach of the ISO to Jupiter and the Juno spacecraft on Mar 16, 2026. 

%% Please use the acknowledgment and contribution environments. This will 
%% be anonomyized when the "anonymous" style option is used. 
\begin{acknowledgments}
We would like to thank Karl Battams, Jim Bell,  Shane Byrne,  Alfred McEwen and Qicheng Zhang
 for useful discussions. T.L. would like to thank Michael Jäger and Gerald Rhemann for providing part of the green filtered images for analysis.
\end{acknowledgments}

\software{findorb 
          \url{https://www.projectpluto.com/find_orb.htm}
          }


%% For this sample we use BibTeX plus aasjournalv7.bst to generate the

\bibliography{3I}{}
\bibliographystyle{aasjournalv7}



\end{document}